\documentclass[aps,prl,showpacs,showkeys,twocolumn,amssymb,groupedaddress]{revtex4}

\usepackage{graphicx}
\usepackage{dcolumn}
\usepackage{bm}

\begin{document}

\title{Universal formula for the muon-induced neutron yield}

\author{N. Yu. Agafonova}
 \email{agafonova@lngs.infn.it}
\author{A. S. Malgin}
 \email{malgin@lngs.infn.it}

\affiliation{Institute For Nuclear Research of the Russian Academy of
Sciences \\
60th October Anniversary Prospect 7a, 117312 Moscow, Russia}

\date{\today}

\begin{abstract}
The experimental data on the yield of muon-induced neutrons for liquid 
scintillator, iron and lead accumulated during 60 years of muon 
interaction underground study have been analyzed. A universal formula 
connecting the yield with muon energy loss in the matter and neutron 
production in hadronic and electromagnetic showers is presented.
\end{abstract}

\pacs{25.30.Mr}

\keywords{neutron yield, muons}

\maketitle

{\em Intriduction.-}In the last decade a renewed interest to the problem of yield $Y_n$ 
of muon-induced neutrons becomes evident. This is due to both the 
increased requirements to the accuracy of background definition in 
underground experiments and the growth of computing resources. The 
yield dependence on both the mean muon energy ${\overline E}_{\mu}$ and atomic 
weight $A$ of the medium has been investigating using the FLUKA and GEANT 
simulation packages and their versions  \cite{Wa01}, \cite{Ar05}, 
\cite{Me06}, \cite{Ma08}. The yield calculation results are significantly different 
(Fig.1 in \cite{Ar05}). The normalization of calculations to the experimental 
data becomes more complicated due to large straggling in these data. 
Until now no expression exists for $Y_n$ that binds together the muon 
energy deposition, nuclear properties of the matter and the neutron 
production processes in hadronic (h) and electromagnetic (em) showers 
generated by muons and developing in the matter. To calculate a yield 
value the approximate empirical laws 
$Y_n=p_E{\times}E^{\alpha}_{\mu}$ (for fixed $A$) or 
$Y_n=p_A{\times}A^{\beta}$ 
(for fixed $E_{\mu}$) are used. The constants ${\alpha}$, ${\beta}$ 
are defined based on the results of calculations. Numerical fitting 
coefficients $p_E$, $p_A$ are entered to get the agreement between 
calculations and a set of experimental data available.

The form of dependence $Y_n({\overline E}_{\mu})= a{\overline E}^{\alpha}_{\mu}$ 
was proposed in \cite{Za65}. 
As follows from results of the measurements \cite{Bez73}, \cite{En87}, 
\cite{Ag89}, \cite{Go71} and calculations \cite{Wa01}, \cite{Ar05}, 
\cite{Za65}, \cite{Gr72}, \cite{Pe11}, the ${\alpha}$ value 
is in the range from 0.7 to 0.9. The values of exponents  ${\alpha}$ 
and ${\beta}$ representing the contributions of the neutron 
production channels play an important role in the analysis of the 
experimental and calculated data.

Experimentally, the yield $Y_n$ is given by  \\
\begin{equation}
Y_n=\frac{N_n}{{\overline l}_{\mu}{\rho}}(n/g/cm^2),  \label{eq1}\\
\end{equation}

where $N_n$ is the number of neutrons produced by a muon at energy 
${\overline E}_{\mu}$
on the path length ${\overline l}_{\mu}$ in the matter with density ${\rho}$.
 ${\overline E}_{\mu}$ and  ${\overline l} _{\mu}$ are mean values 
for muon flux at a depth $H$  
and muon track lengths, respectively. The yield is 
connected with the medium properties and the 
characteristics of reactions of neutron production by expression: \\
\begin{equation}
Y_n=\frac{N_0{\langle}{\nu}{\sigma}{\rangle}}{A}, \label{eq2}\\
\end{equation}

where $N_0$ is the Avogadro number, ${\langle}{\nu}{\sigma}{\rangle}$ is a mean 
value of the product of photonuclear ${\mu}A$-interaction cross 
section and neutron multiplicity ${\nu}$, $A$ is the atomic weight of the medium.
Equation (\ref{eq2}) follows from the dependence $N_n$ on ${\langle}{\nu}{\sigma}{\rangle}$ 
and ${\overline l}_{\mu}$\\
\begin{equation}
N_n=c_A{\langle}{\nu}{\sigma}{\rangle}l_{\mu}=\frac{{\rho}N_0}{A}{\langle}{\nu}{\sigma}{\rangle}=\frac{{\langle}{\nu}{\sigma}{\rangle}}{A}{\rho}l_{\mu}N_0, \label{eq3}\\
\end{equation}
where  $c_A$ $[cm^{-3}]$ is a concentration of nuclei $A$. \\

{\em Experimental data.-}The table lists the measured yield $Y_{LS}$ for liquid scintillator 
(LS), Fe ($Y_{Fe}$), Pb ($Y_{Pb}$). 
The table covers almost all the data obtained for about 60 - year 
measurements of the neutron yield in underground experiments. 
The data are listed in the order of increasing of energy
${\overline E}_{\mu}$ to which authors attributed their result.
An error in determining average muon energy ${\overline E}_{\mu}$ was only 
shown in Ref. \cite{Ab10}. To estimate the  ${\overline E}_{\mu}$ value error 
in other experiments summarized in the table we have used the expression 
${\delta}{\overline E}_{\mu}= 2/{\sqrt {{\overline E}_{\mu}}}$. It covers both the 
uncertainty ${\overline E}_{\mu}$ calculations at
different sets of parameters offered in Ref. \cite{Gu84}, \cite{Li91},\cite{Do01} and
deviations of ${\overline E}_{\mu}$ values from the ${\overline E}_{\mu}(H)$ 
dependence which can be seen in the  table.
The Mote Carlo calculations carried out recently \cite{Zb10},
\cite{Ag13} have resulted in a revision some $Y_{LS}$ values. 
The most part of the LS data \cite{Bez73}, \cite{Ag89}, \cite{Pe11}, \cite{He95}, \cite{Bo00},
 \cite{Ry86}, \cite{Ag05} was obtained using 
scintillator $C_nH_{2n}$ $n=9.6$ ${\rho}=0.78 g/cm^3$ 
\cite{Bez73}, \cite{Pe11}, \cite{Ry86}, \cite{Ag05} 
(in the table the refined value $Y_{LS} = 4.1{\times}10^{-4} n/{\mu}/(g/cm^2)$ 
from \cite{Ag11} is included).

It should be noted that the paper \cite{Me06} cites wrongly (Table IV in \cite{Me06})
the results of Ref. \cite{Bez73}, \cite{Ag89}, \cite{Ry86},
namely, out of 15 values taken from these works and included in the
Table IV, seven values do not correspond to the published original
data. The correct values of $H$,
${\overline E}_{\mu}$ and $Y_{LS}$ are presented in the table of given paper
and also in Ref. \cite{Ma08}.

\begin{widetext}
\begin{center}
\begin{table}[t]
\centering
\caption{\label{tab:data}Measured neutron yield}
\begin{ruledtabular}
\begin{tabular}{@{}l*{15}{l}}

 &  &  \multicolumn{3}{c}{\textbf{$Y_n{\times}10^{-4}, n/{\mu}/(g/cm^2)$}} &   & \\
\textbf{${\overline E}_{\mu}, GeV$}        & \textbf{H, m.w.e.} & \textbf{$Y_{LS}$} & \textbf{$Y_{Fe}$} & \textbf{$Y_{Pb}$ } & \textbf{Ref.} & \textbf{year} \\
\hline
$10.0{\pm}6.3$\footnotemark[1] &      20            &     -           &$0.98{\pm}0.01$  & $2.43{\pm}0.13$  &   \cite{An54}       &  1954         \\
$10.0{\pm}6.3$\footnotemark[1] &      60            &     -           &     -           & $4.8{\pm}0.6$     &  \cite{Be70}       &  1970         \\
$11.0{\pm}6.6$\footnotemark[1] &      40            &     -           &$1.32{\pm}0.30$  & $4.03{\pm}0.36$   &  \cite{Go71}       &  1971         \\
$13.0{\pm}7.2$                 &      20            & $0.20{\pm}0.07$   &     -           &     -           &  \cite{He95}       &  1995         \\
$16.5{\pm}8.1$                 &      32            & $0.36{\pm}0.03$   &     -           &     -           &  \cite{Bo00}       &  2000         \\
$16.7{\pm}8.2$                 &      25            & $0.47{\pm}0.05$   &     -           &     -           &  \cite{Bez73}       &  1973         \\
                             &                    & $0.36{\pm}0.05$\footnotemark[2]&    &                 &               &               \\
$17.8{\pm}8.4$\footnotemark[1] &      80            &     -           &$1.69{\pm}0.30$  & $5.66{\pm}0.36$   &  \cite{Go71}       &  1971         \\ 
$20{\pm}9$\footnotemark[1]     &     110            &     -           &     -           & $6.8{\pm}0.9$     &  \cite{Be70}       &  1970         \\
$40{\pm}12.6$\footnotemark[1]  &     150            &     -           &$3.31{\pm}0.96$  & $11.56{\pm}1.1$   &  \cite{Go68}       &  1968         \\
$86{\pm}18$                    &     316            & $1.21{\pm}0.12$   &     -           &     -           &  \cite{Bez73}       &  1973         \\
                             &                    & $0.93{\pm}0.12$\footnotemark[2]&    &                 &               &               \\
$110{\pm}21$\footnotemark[1]   &     800            &     -           &     -           & $17.5{\pm}3.0$    &  \cite{Go70}       &  1970         \\
$125{\pm}22$                   &     570            & $2.04{\pm}0.24$   &     -           &     -           &  \cite{Ry86}       &  1986         \\
                             &                    & $1.57{\pm}0.24$\footnotemark[2]&    &                 &               &               \\
$260{\pm}8$                    &    2700            & $2.8{\pm}0.3$     &     -           &     -           &  \cite{Ab10}       &  2010         \\
$280{\pm}33$                   &    4300            &     -           &     -           & $116{\pm}44$      &  \cite{Be73}       &  1973         \\
$280{\pm}33$                   &    3100            & $4.1{\pm}0.5$     &  $16.4{\pm}2.3$   &     -           &  \cite{Ag05}       &  2005         \\
                             &                    & $3.3{\pm}0.5$\footnotemark[2]&      &                 &               &               \\
$280{\pm}33$                   &    3100            & $3.2{\pm}0.2$     &  $19.0{\pm}1.0$   &     -           &  \cite{Pe11}       &  2011         \\
$385{\pm}39$                   &    5200            & $5.3^{+0.95}_{-1.02}$   &  $20.3{\pm}2.6$   &     -           &  \cite{Ag89}       &  1989         \\
                             &                    & $4.1{\pm}0.6$\footnotemark[2]&      &                 &               &               \\
\end{tabular}
\end{ruledtabular}
\footnotetext[1]{Vertical flux}
\footnotetext[2]{Corrected values}
\end{table}
\end{center}
\end{widetext}

The measurements were carried out in a global muon flux at different
depths and energies ${\overline E}_{\mu}$ from 16.7 GeV to 385 GeV. The
experiments \cite{Bez73}, \cite{Ry86} detected the neutrons produced only in
the counter LS; the
results of Ref. \cite{Ag89}, \cite{Pe11}, \cite{Ag05} covered the
neutrons generated in LS and iron of the setup structures (LS and
iron masses were almost equal). 

The counters were located close to the
mine ceiling of gypsum (${\overline E}_{\mu}$ = 16.7 GeV) or salt
(${\overline E}_{\mu}$= 86 GeV) in the experiment \cite{Bez73}
and close to the mine ceiling of salt (${\overline E}_{\mu}$ = 125 GeV)
in Ref. \cite{Ry86}. As a result of the Monte Carlo calculations in
Ref. \cite{Zb10}, it was obtained that the contribution
of neutrons produced by shower particles in the standard rock around
the detecting volume LS ($C_{12}H_{26}$) enlarges the measured
yield $Y_{LS}$ by ${\sim} 30\% $. Taking into account this fact and
disregarding the small difference between compositions of LS and
rock in experiments \cite{Bez73}, \cite{Ry86} and calculations \cite{Zb10},
we have obtained the corrected values of $Y_{LS}$ which are presented
in the table.

The similar correction is not suited for the results of 
\cite{Ag89}, \cite{Ag05}, \cite{Ag11} since in these 
experiments inner counters are detecting the neutrons produced in an 
inner volume of setup consisting of LS and iron in the same proportion 
as the peripheral part of the setup.

With LSD and LVD which are almost identical in design and detection 
technique the yield was measured under different conditions of 
neutron detection: a) with inner counters crossed by a muon 
(LSD \cite{Ag89}), b) with all counters of the inner setup volume crossed 
by a muon (LVD \cite{Ag05}, \cite{Ag11}), c) with inner counters fired by any 
trigger pulse, including the muon trigger (LVD \cite{Ag99}). Here a muon 
is meant both a single muon and a muon group with shower accompaniment 
or without it. In all papers using the LSD and LVD data the yield was 
defined by formula\\

\begin{equation}
Y_{LS}=\frac{N^{det}}{N_{\mu}{\rho}_{LS}{\overline l}_{LS}{\eta}}, \label{eq4}\\
\end{equation}

where $N^{det}=N^{det}_{LS}+N^{det}_{Fe}$  is the number of detected 
neutrons, including produced in LS ($N_{LS}$) and iron ($N_{Fe}$), 
while $N^{det} = N_{LS}{\eta}_{LS} + N_{Fe}{\eta}_{Fe}$, 
where ${\eta}_{LS}$, ${\eta}_{Fe}$ are corresponding neutron detection 
efficiencies; $Q$ is the fraction of neutrons produced in LS, 
$N_{\mu}$ is an amount of muons, ${\overline l}_{LS}$ is a mean length of muon tracks 
in LS. $N^{det}, N_{\mu}$ and ${\overline l}_{LS}$ have been determined directly 
in the experiment. $N_{\mu}$ has been determined with due regard to the 
multiplicity of muon groups established using the tracking system data 
\cite{Ag05}. The Q fraction was calculated with the assumption that 
${\eta} = {\eta}_{LS} = {\eta}_{Fe}$. The values of $Q$, 0.61, 0.60, 0.85 
and ${\eta}$ = 0.60, 0.90, 0.60 have been used for cases a), b) c), 
respectively. Case c) leads to the selection of neutrons at energies 
above 10 MeV and, as a consequence, to significant reduction in 
$Y_{LS}$ what was indicated in \cite{Ag05}, \cite{Ku03}. For this reason, the 
result of \cite{Ag99} is not included in the table.\\

The recent Monte Carlo calculations in Ref. \cite{Ag13} showed that 
${\eta}_{LS} {\neq} {\eta}_{Fe}$. This leads to the need to change 
the formula (\ref{eq4}):\\

\begin{equation}
Y_{LS}=\frac{N^{det}}{N_{\mu}{\rho}_{LS}{\overline l}_{LS}}{\times}\frac{Q}{Q{\eta}_{LS}+(1-Q){\eta}_{Fe}}. \label{eq5}\\
\end{equation}

Given fraction $Q$ the yield $Y_{Fe}$ can be also defined\\

\begin{equation}
Y_{Fe}=\frac{N^{det}}{N_{\mu}{\rho}_{Fe}{\overline l}_{Fe}}{\times}\frac{1-Q}{Q{\eta}_{LS}+(1-Q){\eta}_{Fe}}, \label{eq6}\\
\end{equation}

where ${\overline l}_{Fe}$  is a mean length of muon tracks in iron and 
${\rho}_{Fe}$ is the iron density.\\
New $Q$ values were calculated in \cite{Ma12} for cases a), b) of the 
neutron detection with LSD and LVD. The $Q$ fraction depends on the 
ratios of masses $k_M=M_{LS}/M_{Fe}$, surface areas 
$k_S=S_{Fe}/S_{LS}$ calculated per counter, atomic weights 
$k_A = A_{LS}/A_{Fe}$, and the exponent ${\beta}$\\
\begin{equation}
Q = \frac{k^{\beta}_Ak_Mk_S}{(1+k^{\beta}_Ak_Mk_S)}. \label{eq7}\\
\end{equation}

\begin{figure}[!t]
\centering
\includegraphics[width=2.8in]{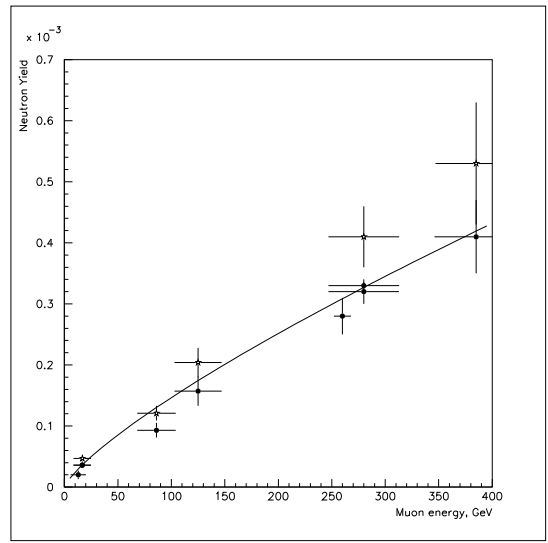}
\caption{Dependence of the neutron yield on muon energy for scintillator. 
The curve is a function $Y_n=4.03{\times}10^{-6}{\overline E}_{\mu}^{0.78}$ 
fitting the experimental points (filled circles), open stars are uncorected data.}
\label{1fig}
\end{figure}

On the basis of the LSD experimental data in \cite{Ma12} it was found 
that ${\beta}$ = 0.95 and $Y_{Fe} = 20.3{\times}10^{-4}$. 
To do this, the data of inner counters of first LSD level were 
used. In addition, these counters were detecting the neutrons 
produced by muons in a 8-cm thick steel platform beneath the setup. 
Using the data of the inner LSD counters of the second level, the 
fraction $Q = 0.138$ has been determined for case a). The $Y_{LS}$ value 
of $(4.1{\pm}0.6){\times}10^{-4}$ corresponds to this fraction at 
efficiencies ${\eta}_{LS}$ = 0.45, ${\eta}_{Fe}$ = 0.10 
and ${\beta}=0.95$. In the LVD experiment the value $Q$ = 0.18 and 
the corresponding yields 
$Y_{LS}=(3.3{\pm}0.5){\times}10^{-4}$, 
$Y_{Fe} = (16.4{\pm}2.3){\times}10^{-4}$ were 
obtained under detection conditions b). Efficiencies 
${\eta}_{LS}$ = 0.75, ${\eta}_{Fe} = 0.65 $ were taken from \cite{Ag13}. 
Thus, the $Y_{LS}$ values from the reviewed papers exceed the 
corrected magnitudes by $\sim$ 30 \% (the table, Fig.~\ref{1fig}).\\
The recent LVD results have been presented in \cite{Pe11}: 
$Y_{LS} = (3.2{\pm}0.2){\times}10^{-4}, 
Y_{Fe} = (19{\pm}1){\times}10^{-4}$. The yield values were obtained 
based on the data of counters without triggering pulses to avoid 
some methodical effect reducing neutron detection efficiency in a 
counter with triggering pulse. In this case the efficiencies 
are ${\eta}_{LS} = 0.0075, {\eta}_{Fe} = 0.0107$. All quantities 
($Q$, ${\eta}$, ${\overline l}$) except for the starting number 
$N^{det}$ of the 
detected neutrons were calculated by the Monte Carlo method.

\begin{figure}[!t]
 \centering
 \includegraphics[width=2.8in]{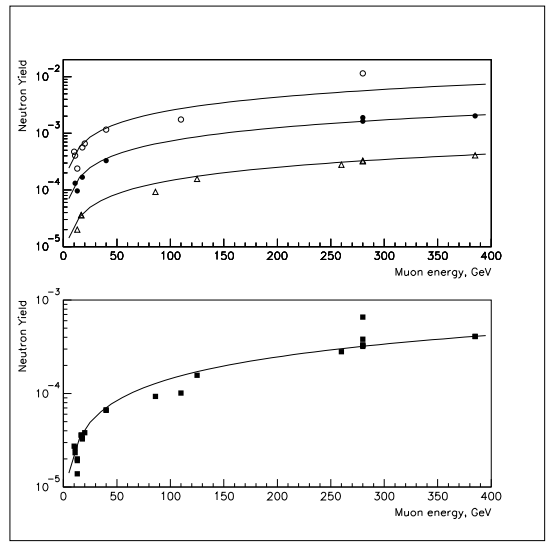}
 \caption{Dependence of the neutron yield on $A$ and ${\overline E}_{\mu}$. 
Upper panel: experimental points for lead (A=207, open circles),  
iron (A=56, filled circles) and 
scintillator (A=10.3, open triangles);
the curves are functions $Y=cA^{\beta}{\overline E}_{\mu}$
at different $A$ and  $c=4.4{\times}10^{-7}, {\beta} = 0.95, {\alpha} = 0.78$. 
Lower panel: neutron yield for scintillator; the experimental 
data for iron and lead are reduced to scintillator, 
the curve is a function $Y_{LS}=4.4{\times}10^{-7}10.3^{0.95}{\overline E}_{\mu}^{0.78}$}
 \label{2fig}
\end{figure}

{\em Formula for the muon-induced neutron yield.-} The data 
collected in the table including early measurements \cite{An54}, 
\cite{Be70}, \cite{Go68}, \cite{Go70}, \cite{Be73} with iron and lead were 
analyzed using the conventional approach: ${\alpha}$ and  ${\beta}$ are 
constants independent of  ${\overline E}_{\mu}$ and $A$, correspondingly.
Using independence of  ${\alpha}$ on ${\overline E}_{\mu}$ , for any A we can 
reduce the yield $Y({\overline E}_{\mu})$ values to a certain arbitrarily 
chosen energy  ${\overline E}^{\ast}_{\mu}$ and calculate the average value of 
${\langle}Y({\overline E}_{\mu}){\rangle}$: for ${\overline E}^{\ast}_{\mu}$ = 16.7 GeV ${\langle}Y_{LS}{\rangle}$ = 0.34 
(averaged over nine values),  ${\langle}Y_{Fe}{\rangle}$     
= 1.70 (averaged over seven values), 
${\langle}Y_{Pb}{\rangle} = 6.33{\times}10^{-4} n/{\mu}/(g/cm^2)$ (averaged over 
eight values).The ratio ${\langle}Y_{LS}{\rangle}/{\langle}Y_{Fe}{\rangle}$ is consistent with 
${\beta}$ = 0.95, while ${\langle}Y_{LS}{\rangle}/{\langle}Y_{Pb}{\rangle}$ with  ${\beta}$ = 0.97 
and ${\langle}Y_{Fe}{\rangle}/{\langle}Y_{Pb}{\rangle}$ with ${\beta}$ = 1.00. The large  ${\beta}$ values 
in the last two cases are mostly associated with excessive yield 
$Y_{Pb} = 116{\times}10^{-4}$ in experiment \cite{Be73}.\\  
The table data presented in Fig.~\ref{2fig}  can be described by the expression\\

\begin{equation}
Y_n(A,{\overline E}_{\mu}) = cA^{\beta}{\overline E}_{\mu}^{\alpha}, \label{eq8}\\
\end{equation}

where ${\beta}$ = 0.95, c is constant.
Using the independence of ${\beta}$  on A and assuming ${\beta}$ = 0.95, 
the $Y_{Fe}({\overline E}_{\mu})$ and $Y_{Pb}({\overline E}_{\mu})$ data sets can be reduced 
to the $Y_{LS}({\overline E}_{\mu})$ set (Fig.~\ref{2fig}, lower panel). Fitting the yield set of 24 values 
$Y_{LS}({\overline E}_{\mu})$ by expression $Y_{LS} = c(10.3)^{0.95}{\overline E}_{\mu}^{\alpha}$
we get the best agreement with the data at 
$c = (4.4{\pm}0.3){\times}10^{-7}$ and ${\alpha} = 0.78{\pm}0.02$. 
The same values $c$ and ${\alpha}$, but at larger uncertainties, 
result from corrected LS data (Fig.~\ref{1fig}, nine values). \\
The constant c is close to the value of the relative muon energy 
loss in nuclear interactions $b_h = 4.0{\times}10^{-7} (g/cm^2)^{-1}$. 
Therefore, $c$ is a relative muon energy loss for neutron production 
$c = b_n$ and it has the dimension $(g/cm^2)^{-1}$. Since neutrons are produced 
mainly in the em- and h- showers the constant $b_n$ should be associated 
with electromagnetic energy loss of muAgafonova2.texons $b_{\gamma}$ (em-shower 
generation mostly by means of  bremsstrahlung) and nuclear loss 
$b_h$ (generation of h-showers where the main part of neutrons is produced). 
In the range from ${\sim}$ 100 GeV to the extreme average muon energy 
underground ${\sim}$ 430 GeV the generation of em- and 
h-showers is proportional to ${\overline E}_{\mu}$. The $b_h$ 
value does not depend on ${\overline E}_{\mu}$ and weakly depends on A: 
$b_h = 4.0{\times}10^{-7}$ for standard rock, $4.2{\times}10^{-7}$ 
for water \cite{Gu84}; the $b_{\gamma}$ loss varies slightly 
from $12.1{\times}10^{-7}$ up to $14.2{\times}10^{-7}$  for rock and 
from $8.2{\times}10^{-7}$ up to $9.9{\times}10^{-7}$  for water \cite{Gu84}. 
The $b_{\gamma}$ loss depend on the medium as $Z^2/A$. 
The proximity of the values $b_n$ and $b_h$ reflects, on one hand, 
the dominant role of h-showers in neutron production and, on the other, 
the practical constancy of $b_n$ in a wide range of ${\overline E}_{\mu}$ 
and $A$.
The values of exponents ${\alpha}$,${\beta}$  in equation (\ref{eq8}) are 
determined by neutron production processes in showers: in em- showers 
$Y_n {\propto} {\overline E}_{\mu}^{1.0}$ \cite{En87}, 
in h-showers $Y_n {\propto} {\overline E}_{\mu}^{0.75}$ 
\cite{Ag89}, \cite{Gr72}, \cite{Ry86}. Therefore, the resultant 
values ${\alpha}$ = 0.78, ${\beta}$ = 0.95 and 
$b_n = 4.4{\times}10^{-7}$ obtained above are associated with 
the contributions of all neutron production processes, namely the 
shower generation by muons and the neutron production in showers 
via ${\pi}A$, $NA$, ${\gamma}A$-reactions.\\
Since the product $b_nE_{\mu}^{\alpha}$  $[GeV/{\mu}/(g/cm^2)]$ defines 
the muon energy loss for the neutron production, then, due to the 
yield dimension $[n/{\mu}/(g/cm^2)]$, factor $A^{\beta}$ 
has the dimension of [$n/GeV$].\\ 
The yield value is contained in the formula for the neutron 
production rate
 $r_n=I_{\mu}(H){\rho}_AY_n(E_{\mu},A)$ $(n/cm^3 c)$,\\  
where $I_{\mu}(H)$ $({\mu}/cm^2 c)$ is a muon intensity at a depth $H$ 
and ${\rho}_A$ is a medium density.
Using this formula one can write the expression for rate $R_n$ of 
muon-induced neutrons in the detector and its shield consisting of 
different materials. The neutron rate for a material $A_i$ of a 
volume $v_i$ and mass $m_i$ is given by\\
\begin{equation}
\label{eq9}
R_{ni}=v_ir_n=I_{\mu}(H){\rho}_{A_i}v_iY_{n_i}=I_{\mu}(H)m_iY_{n_i} (n/c).
\end{equation}
For all materials of the detector and shield we have\\
\begin{equation}
\label{eq10}
R_n=I_{\mu}(H){\Sigma}m_iY_{n_i}=I_{\mu}(H)b_nE_{\mu}^{\alpha}{\Sigma}m_iA_i^{\beta} (n/c).
\end{equation}
As it follows from equation (\ref{eq8}) the neutron yield is highly dependent on 
${\overline E}_{\mu} ({\propto}{\overline E}_{\mu}^{0.78})$ and 
$A ({\propto}A^{0.95})$. So,  its value for the heavy 
material (Fe, Pb) can be used for 
experimental determination of ${\overline E}_{\mu}$ at 
any overburden topography and rock composition.
The accuracy of the procedure might be not worse than in finding ${\overline E}_{\mu}$
 by formulae in Ref. \cite{Gu84}, \cite{Li91}, \cite{Do01}. An approximation with constant parameters 
$b_n=4.4 {\times} 10^{-7} cm^2/g, {\alpha} = 0.78, {\beta} = 0.95$
allows to use the formula (\ref{eq8}) to calculate the yield for
 any  ${\overline E}_{\mu}$ and $A$ in underground 
experiments.
Since all nuclear effects produced by muons in the matter, including the
production of radionuclides, are proportional to the neutron yield value
the formula (\ref{eq8}) is universal. 
However, the magnitudes of the parameters are 
determined by the contributions of nuclear and electromagnetic 
processes and therefore albeit weakly but depend on 
${\overline E}_{\mu}$ and $A$. Due to the increasing requirements to the accuracy 
of the background determination in underground experiments the study of the neutron 
yield remains actual.\\
This work was supported in part by the Russian Foundation for
Basic Research grants 12-02-00213-a,12-02-12127-ofi-m and SSh-871.2012.2.

\end{document}